# Multimodal Chaptering for Long-Form TV Newscast Video

Khalil Guetari[1], Yannis Tevissen[1], and Frédéric Petitpont[1]

[1] Moments Lab Research, 92012 Boulogne-Billancourt, France
khalil.guetari@momentslab.com

**Abstract.** We propose a novel approach for automatic chaptering of TV newscast videos, addressing the challenge of structuring and organizing large collections of unsegmented broadcast content. Our method integrates both audio and visual cues through a two-stage process involving frozen neural networks and a trained LSTM network. The first stage extracts essential features from separate modalities, while the LSTM effectively fuses these features to generate accurate segment boundaries. Our proposed model has been evaluated on a diverse dataset comprising over 500 TV newscast videos of an average of 41 minutes gathered from TF1, a French TV channel, with varying lengths and topics. Experimental results demonstrate that this innovative fusion strategy achieves state of the art performance, yielding a high precision rate of 82% at IoU of 90%. Consequently, this approach significantly enhances analysis, indexing and storage capabilities for TV newscast archives, paving the way towards efficient management and utilization of vast audiovisual resources.

## 1 Introduction

Every minute, more than 500 hours of video are uploaded on YouTube. Numbers are even bigger if one considers all social media and broadcasted contents. A big part of these videos are long-form contents such as tutorials, talk-shows, reality-shows or news reportages. Yet, as for today, most video understanding technologies focus on short videos due to compute limitations and temporal coherence necessity.

A paramount pre-processing step in longform video management is about segmenting the video into distinct chapters. Chapters are sequences of semantically coherent segments which are then used for more precise analysis and retrieval [1, 2]. TV newscast videos naturally have this need of segmenting streams of video into logical units for archive and further analysis purposes [3].

This research presents a multimodal approach to segment TV newscasts into chapters based on the fusion of image, audio and text modalities. This approach aims to be generic and do not use specific design rules from the newscast types of videos.

## 2    Related Works

TV newscast chaptering is the task of segmenting a long video into chapters. Chapters are contiguous and non-overlapping segments which completely segment a video [4]. Thus, the end of a chapter is the start of a new chapter. Previous research mentions this task as story segmentation for TV news which was often evaluated on the TRECVID 2003 dataset [5]. Stories are defined as "segments of a news broadcast with a coherent news focus which contain at least two independent declarative clauses" [6].

A common approach used for TV news story segmentation is by detecting the anchor person [6, 7, 8]. Hmayda et al. [7] uses optical flow and a face detection model to have a list of potential anchor persons. Candidate anchor persons are then clustered and the cluster with the highest number of candidates is picked as the anchor person. Other visual and audio descriptors like text screen, logo, silence and transition words were fused by Dumont et al. in a trained classifier [8].

In addition to the assumption of the presence of an anchor person at the beginning and the end of a TV newscast, some research often relies on the use of specific TV newscasts' design rules like word repetition [9] or cue phrases [10]. Kannao et al. [11] uses text similarity between overlays correlated with Web news articles to classify shots in four categories describing the stories segments boundaries. Text similarity was used more recently by training an LSTM based Siamese neural network on sentences belonging to the same and different stories (positive and negative pairs respectively), in the context of Urdu news [1].

TV newscast chaptering is also related to the task of video scene segmentation where unsupervised learning was mostly used in previous approaches. Shots are represented in a feature space using low level features and several clustering techniques are used to group shots in scenes [12, 13]. More recent techniques use deep learning to fuse different modalities into a trained neural network. Rotman et al. [14] uses visual and audio embeddings from frozen backbones. Embeddings are then trained in a fully connected network and fused together. Tiago et al. [15] trains an LSTM to output if a shot is a shot transition between two scenes. CSIFT visual features [16] and MFCC features [17] are passed through a CNN each before being fed to the LSTM network.

In this research, we use a bidirectional LSTM to fuse visual, textual and audio features from frozen backbones. Identified speakers and silences are also fused in this architecture. The LSTM is trained to output a distance matrix used to create chapters. We conduct several experiments on different fusion block architectures and show we obtain best results on a bidirectional LSTM architecture.

## 3    Proposed Method

### 3.1    Overview

Our approach is based on a two-stage process. The first step is to extract features from the different modalities of the raw video. Image embeddings and text embeddings are retrieved using pretrained models. Speakers and silences are identified and encoded. Audio features are extracted by using the MFCC algorithm [17].

The second step is to fuse extracted modalities into a trained Bidirectional LSTM (Bi-LSTM) which returns processed features. A distance matrix is generated and is used to detect chapters' boundaries. The training is done by using a Block Adjacent Froebenius Loss driving the Bi-LSTM into maximizing the distances between shots of distinct chapters.

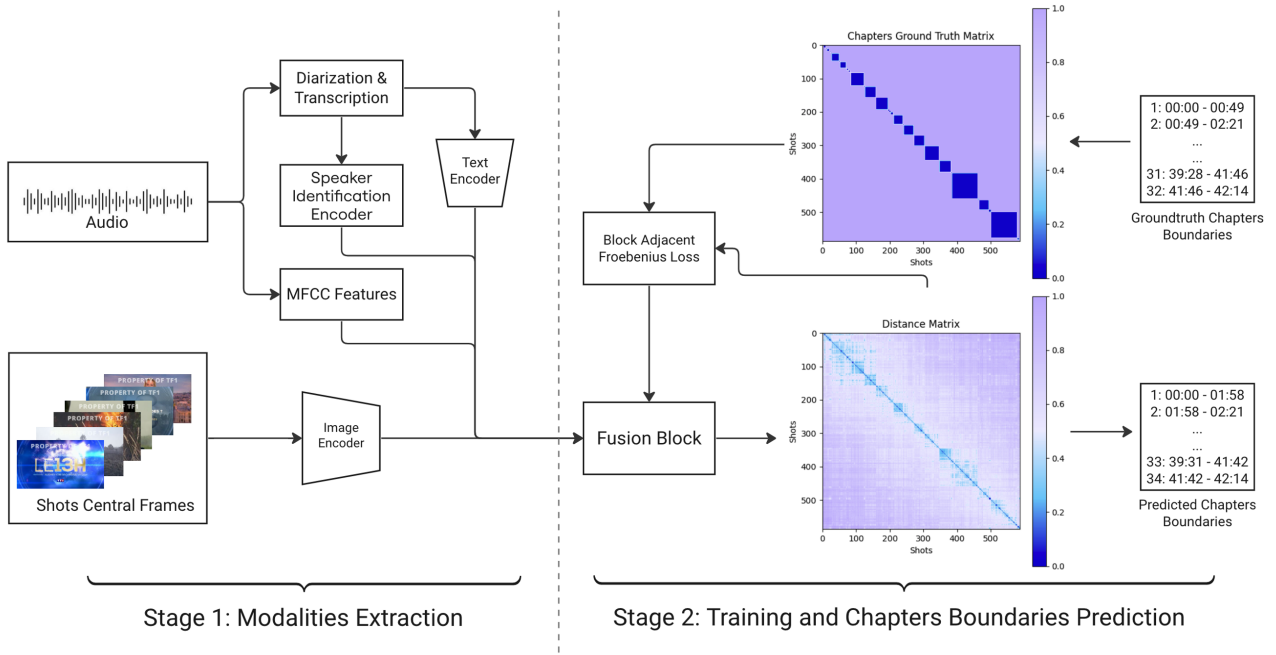

**Fig. 1.** Overview of the two-stage process to detect chapters. The first stage extracts features for every shot of the newscast. Visual and text features are extracted from frozen image and text encoders. MFCC features are extracted, and speakers are identified. The second stage fuses features in a fusion block. This block is trained by computing the Block Adjacent Froebenius Loss between predicted and ground truth distance matrixes.

### 3.2  Chaptering Data

A collection of 531 chaptered TV newscast videos was collected for this research. This collection comes from TF1, one of the top most viewed French television channels.

For each video, chapters were annotated by a documentalist in an Excel spreadsheet specifying start and end timestamps of each chapter. Chapters consist of news stories, in-studio interviews and presenter story's introductions. The average duration of a video is 41 minutes, and the average number of chapters is 35. Figure 2 shows histograms for the distribution of number of shots, chapters and the duration of video.

To be more compute efficient, we decide to model the chapters detection as a shot boundary detection. This enables it to process fewer frames. Moreover, TV newscast stories are often separated by a distinct visual change between shots. Shots are detected using Pyscenedetect with a custom shot detector. This custom shot detector enables it to deal with the numerous fading transitions in TV newscast video. The default behavior of Pyscenedetect is to detect shots based on the difference between each frame's visual descriptors (hue, saturation, and brightness values). The custom detector compares the frame's visual descriptors with a temporal window of ten frames giving the possibility to detect fades transition in the video.

Speaker diarization is performed with pyannote 2.1 [18] with a hard limit of 20 speakers per file. A third-party API (Vocapia) is used to transcribe audio segments with French as the target language. Timestamps at word level are also returned.

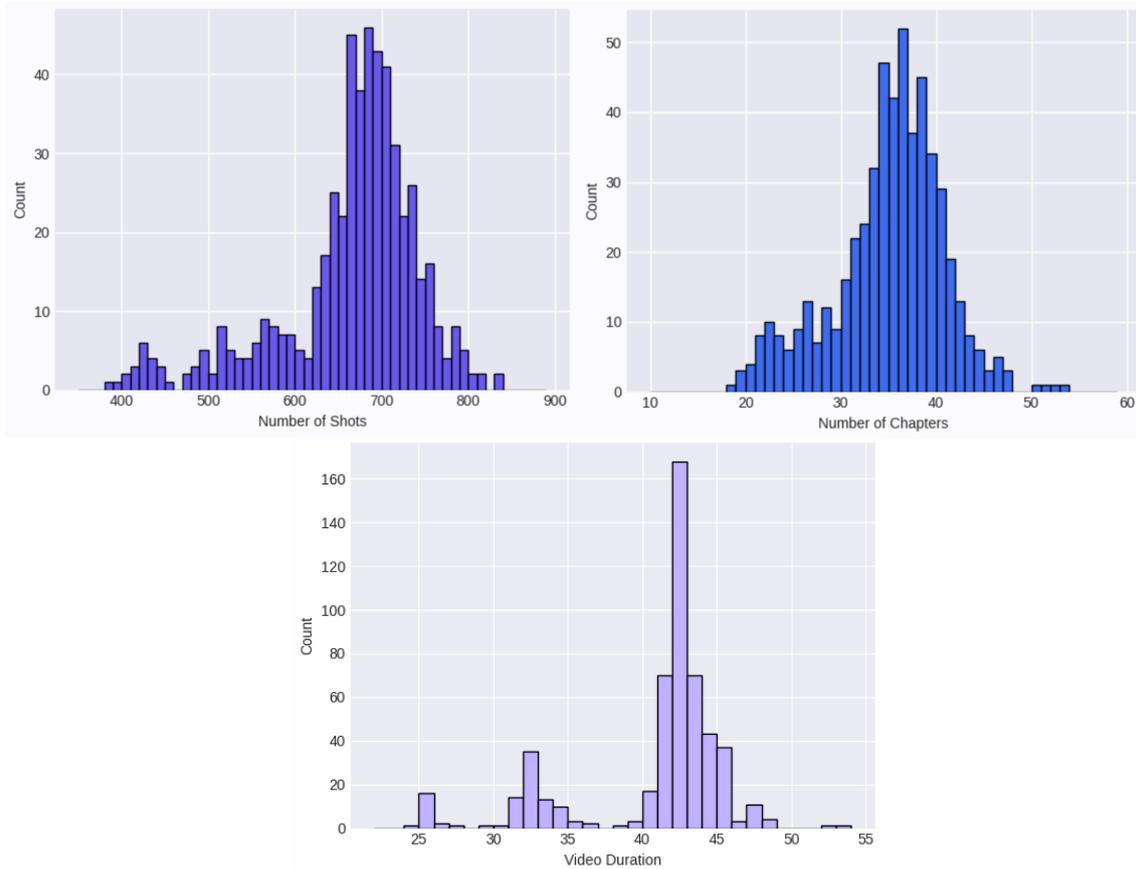

**Fig. 2.** Statistics of the TF1 TV Newscasts Dataset

### 3.3   Audio and Visual Features

Four modalities are extracted from the raw video and are aligned with the detected shots. Features for each modality are extracted, concatenated and stored offline to be more efficient during the training phase.

**Visual Modality.** Image embeddings are extracted from a frozen Large Vision Transformer pretrained on ImageNet [19, 20]. The classification token is extracted to model the visual feature of the shot with a dimension of 1024. The central frame of the shot is used.

**Speaker Modality.** Speakers in the video are identified during the diarization step. There may be several speakers in the same shot so a simple rule to associate a speaker to a shot is to take the speaker with the maximum time of speech in the shot. Potential silences are also associated with a shot when no speaker is found. Speakers and silences are one-hot encoded. This modality will help the model in understanding the patterns linked to the story news and the news anchor.

**Textual Modality.** Transcript segments are aligned with detected shots using timestamps. Since the spoken language is in French, we use the known French trained language model CamemBERT [21]. To avoid having small chunks of words with not enough context, the text is not shot based but the whole transcript segment is given to a shot.

**Audio Features.** Audio is extracted using FFMPEG. From the audio, MFCC features [17] are extracted using Librosa library. Audio features are aligned with shot timestamps by averaging the different MFCC values for a given shot. This gives an audio representation of a shot.

### 3.4   Modalities Fusion Model

Several architectures have been tested to fuse the input features. Following [14], we test with a simple Dense Neural Network (DNN).

Taking inspiration from more complex architectures used for natural language processing tasks, we also use an LSTM [22] and a Transformer architecture [23]. A video segmented by shots can be seen as a sequence of tokens. A token models the embedded features for one shot and a sequence of these tokens are fed to the network.

These architectures are used to extract relevant features from the fused modalities. Based on the extracted features, a matrix of distances between the shots of the video is generated. Thus, a fully connected head layer is also added at the end of LSTM and Transformer models to generate distances matrixes.

Different experiments were conducted on these models with several hyperparameters values. Results are shown in section 4.

### 3.5 Block Adjacent Frobenius Loss

For each shot of the input video, a feature is extracted from the model trained on the fusion of modalities and on the chapters to detect. Following the idea of [14], these extracted features are then compared one to another to create a matrix of distances. Using a threshold, this matrix is used to create the output chapters. If the distance between two shots is above the threshold, then a chapter boundary is set at this location. Thus, the loss function of this approach is defined and then computed by comparing the output matrix and the ground truth matrix. This loss function takes inspiration from the Frobenius Loss and is called Block Adjacent Frobenius Loss. $D$ is the predicted distance matrix and $D^*$ is the target matrix. An example of a predicted matrix and its corresponding target matrix can be seen in figure 1. This loss is block adjacent, meaning that we consider only adjacent chapters to compute the loss. Equation 1 defines the set of shots that are located in adjacent chapters. $x_j$ represents the feature vector extracted from the shot $j$. $L$ is a function to associate shots and their corresponding chapters indexes.

$$A = \{i, j \in \mathbb{N}^2, L(x_i) - 1 \leq L(x_j) \leq L(x_i) + 1\} \tag{1}$$

Equation 2 defines the Block Adjacent Frobenius Loss, the norm of the difference of the adjacent elements of the distances matrixes.

$$\|D - D^*\| = \sqrt{\sum_{i,j \, \epsilon \, A} |D_{i,j} - D^*_{i,j}|^2} \tag{2}$$

## 4 Results

In this section, we present results on several experiments. We first compare performances of different fusion models. Previous research, mainly Vid2Seq [24] and an anchor person-based detection [6, 7, 8] are also tested on our benchmark.

### 4.1 Evaluation Metrics

Located chapters are evaluated using precision (P@Ks, P@K) and recall (R@Ks, R@K) across various thresholds for both distance to the ground-truth chapter start timestamp and for the IoU between the predicted and ground-truth start and end timestamps window. F1 score (F1@Ks, F1@K) is also computed. The distance metrics are evaluated at 1, 3 and 5 seconds thresholds and the IoU based metrics are evaluated at 0.5, 0.7 and 0.9 thresholds.

The dataset used was split into training, validation and test sets with the following number of samples respectively: 381, 96, 54.

If not mentioned otherwise, the following results were all evaluated on the test set.

### 4.2 Fusion Model Study

As presented in section 3.4, we compare the trained fusion architectures on our dataset. The architectures giving the best results with their corresponding hyperparameters are the following ones:

- DNN of 4 blocks of fully connected layers with 4000, 3000 and 1000 as the hidden layers' dimensions. Batch normalization [25] and Relu [26] activation function are used.
- Bidirectional LSTM (Bi-LSTM) with a hidden dimension of 128 and a number of 3 layers.
- Transformer with a depth of 4, a number of heads of 12, hidden dimension of 768 and a Gelu [27] activation function.

For each model, dropout [28] are also added. Adam optimizer [29] was the optimizer giving the best results. Training was performed on an Nvidia A10 GPU. Table 1 shows the obtained results. A cosine scheduler was used, the number of epochs varied between 5 to 20 and the batch size was of 1, 2 or 4.

**Table 1.** Evaluation metrics obtained for different modalities fusion architectures.

| Architecture | F1@5s | F1@3s | F1@1s | P@1s | R@1s | F1@0.5 | F1@0.7 | F1@0.9 | P@0.9 | R@0.9 |
|---|---|---|---|---|---|---|---|---|---|---|
| DNN | 89.94 | 85.47 | 79.35 | 73.16 | **87.17** | 89.94 | 77.96 | 69.54 | 64.25 | 76.20 |
| Bi-LSTM | **93.04** | **90.49** | **85.37** | **89.29** | 81.98 | **93.56** | **89.48** | **82.43** | **86.14** | **79.22** |
| Transformer | 86.15 | 82.73 | 77.46 | 71.25 | 85.55 | 81.73 | 74.47 | 66.41 | 61.20 | 73.17 |

### 4.3 Comparison to the state of the art

We compare our approach to Vid2Seq [24], a multimodal single stage transformer-based model trained on VidChapters-7M [4], a dataset of 817K longform videos. We take this model to compare to because it was trained on a dataset of longform videos. In terms of duration, on average, a video lasts 22.6 minutes making it the closest dataset to ours. A first evaluation is done as a zero-shot video chaptering task with version of Vid2Seq pretrained on VidChapters-7M, HowTo100M [30] and on ViTT datasets [31]. We also finetune this model on our training dataset.

We also assess zero-shot capabilities of the first stage of our approach, the multi-modal features extraction, by computing a distance matrix. Lastly, we also compare to an anchor person detection approach. We implement it by extracting image embeddings and by forming clusters of image embeddings. A face detection algorithm enables a better clustering of candidate anchor persons groups. The cluster with the highest number of shots is the selected anchor person cluster. Table 2 show the evaluation metrics obtained for these various experiments.

**Table 2.** Evaluation metrics obtained on state of the art model Vid2Seq. Finetuning on our train dataset is mentioned. Anchor Person refers to the approach of detecting the anchor person in the newscast and Zero Shot (ours) refers to the zero-shot inference based only the extracted features.

| Architecture | Finetuned | F1@5s | F1@3s | F1@1s | P@1s | R@1s | F1@0.5 | F1@0.7 | F1@0.9 | P@0.9 | R@0.9 |
|---|---|---|---|---|---|---|---|---|---|---|---|
| Vid2Seq | No | 6.18 | 3.45 | 1.55 | 4.94 | 0.9 | 11.02 | 3.23 | 0.28 | 0.77 | 0.17 |
| Vid2Seq | Yes | 19.02 | 12.43 | 5.43 | 5.52 | 5.34 | 34.24 | 15.99 | 1.68 | 1.72 | 1.64 |
| Anchor Person | No | 61.18 | 50.64 | 42.66 | 37.53 | 51.58 | 53.15 | 41.45 | 26.97 | 23.59 | 32.78 |
| Zero-Shot (ours) | No | 38.16 | 34.11 | 30.57 | 19.09 | 77.90 | 19.56 | 15.10 | 10.15 | 6.33 | 25.98 |
| Ours | Yes | **93.04** | **90.49** | **85.37** | **89.29** | **81.98** | **93.56** | **89.48** | **82.43** | **86.14** | **79.22** |

## 5  Discussion

### 5.1  Results Interpretation

Comparing the different fusion blocks shows that the Bi-LSTM is the best architecture type. Using both past and future shots to detect chapters makes it suitable for the use of a bidirectional LSTM. The transformer architecture type does not perform well, as expected, transformers-based architecture often needs a big number of data samples to be efficiently trained on. Few fully connected layers (DNN fusion block) show better performances than the transformer architecture giving hints that a simple pattern can be found in the input extracted features.

Regarding the input features, a zero-shot approach based on the direct creation of a distance matrix by using the raw features from the frozen models outperforms Vid2Seq, a generic model trained on a large collection of data. This shows the quality of the features extracted from the frozen models. However, metrics show a high difference between recall and precision values for this method. This is due to the fact of this approach creating too many chapters giving high values of recalls. Segmenting chapters based on the detection of the anchor person also outperforms Vid2Seq and the zero-shot approach proving again the clear pattern that anchor-based approaches rely on.

### 5.2  Limitations

A direct limitation of the Bi-LSTM approach is the impossibility to use this model for live streams content. Even if our use-case is meant for offline processes, it is important to note that this constraint.

Another limitation due to the modeling of chapters detection task as the clustering of shots in a video appears when having video with few and long shots. Examples of these types of videos can appear in broadcasts of press conferences, long interviews or even video podcasts. Since chapters are based on shot boundaries, detected chapters will likely not be accurate anymore. For instance, one would like to create chapters based on topics in a press conference instead of creating chapters based on visual changes. A potential way to solve this limitation would be to use the active speaker detection to segment longer shots into sub-shots. These sub-shots and their associated transcript can then be used to create more precise chapter boundaries.

## 6     Conclusion

The article proposes a novel approach for automatic chaptering of TV newscast videos, addressing the challenge of structuring and organizing large collections of unsegmented broadcast content. The method integrates both audio and visual cues through a two-stage process involving frozen neural networks and a trained bidirectional LSTM network. The experimental results demonstrate that this innovative fusion strategy achieves state of the art performance, yielding a high precision rate of 82% at IoU of 90%, thereby significantly enhancing analysis, indexing and storage capabilities for TV newscast archives.

One promising direction for future work is investigating the use of improved shot detection methods that incorporate speech segments, particularly for handling the limitation of long shots where the transcript plays a crucial role. By refining the shot detection mechanism, the accuracy of chapter boundaries can be enhanced, especially in cases where there is limited visual variation. Additionally, considering the versatility of the proposed approach, it could be interesting to explore its applicability to various genres of long-form videos beyond TV newscasts, including sports events, films, educational videos, and more. Such expansion would enable a comprehensive organization and management of extensive audiovisual repositories, unlocking endless possibilities for researchers, analysts, and enthusiasts alike.

**Acknowledgments.** We thank TF1 for providing access to the data used for this study.